# Study of the Liquid / Vapor Interfacial Properties of Concentrated Polyelectrolyte – Surfactant Mixtures Using Surface Tensiometry and Neutron Reflectometry: Equilibrium, Adsorption Kinetics and Dilational Rheology

Sara Llamas,[1] Eduardo Guzmán,[1,2,*] Andrew Akanno,[1,2] Laura Fernández-Peña,[1] Francisco Ortega,[1,2] Richard A. Campbell,[3] Reinhard Miller[4] and Ramón G. Rubio[1,2]

[1] Departamento de Química Física I-Universidad Complutense de Madrid, Ciudad Universitaria s/n, 28040 Madrid, Spain

[2] Instituto Pluridisciplinar-Universidad Complutense de Madrid, Avda. Juan XXIII, 1, 28040 Madrid, Spain

[3] Institut Laue-Langevin, 71 avenue des Martyrs, CS 20156, 38042 Grenoble, Cedex 9, France

[4] Max-Plank Institute of Colloids & Interfaces, 14475 Potsdam/Golm, Germany

*To whom correspondence should be sent: eduardogs@quim.ucm.es

**Abstract**

The adsorption of mixtures of poly(diallyldimethylammonium chloride) (PDADMAC) and sodium N-lauroyl-N-methyltaurate (SLMT) at the water / vapor interface has been studied using drop profile tensiometry and neutron reflectometry. This study sheds light on the mechanisms involved in the adsorption of polyelectrolyte – oppositely charged surfactants by the characterization of both equilibrium and dynamics features associated with the layer formation at the fluid interface. The results are discussed in terms of an adsorption – equilibration of the interfacial layers as a two-step process: the initial stages involve the adsorption of polyelectrolyte – surfactant complexes formed in the bulk, and a subsequent stage involves reorganization of the interface. This work contributes to the understanding of the physico-chemical features of systems that undergo complex bulk and interfacial interactions with importance in science and technology.

## 1. Introduction

The study of polymer – surfactant mixtures has undergone spectacular growth in recent years, mainly due to their interest in different technological and industrial fields, from drug delivery systems to mineral processing, and from tertiary oil recovery to cosmetic formulations for shampoos and hair care products.[1-8] Among the different existing mixtures that include polymers and surfactants, those containing oppositely charged components present big importance due to their recognized interest in many technological fields.[2, 4] However, there are many issues related to the physico-chemical behavior of such mixtures that remain unclear, especially when their interactions with fluid interfaces are considered.[3, 9-12] The study of polyelectrolyte – surfactant mixtures at the water / vapor interface has been carried out mainly on the basis of surface tension measurements.[13-16] However, in recent years the development of neutron reflectometry has provided access to information related to the structure and composition of such interfaces.[17-19] Despite the extensive studies on the interfacial properties of the aforementioned systems, the scenario is rather complex, mainly due to the fact that polyelectrolyte – surfactant mixtures are strongly affected by non-equilibrium effects. Campbell and Varga have recently provided a detailed picture of the interfacial behavior of polyelectrolyte – oppositely charged surfactant mixtures and its correlations to the bulk phase behavior: depletion of the interface as a result of aggregation in the bulk,[20] and enrichment of the interface as a result of direct interactions of the formed aggregates.[21] This physical picture was based on an understanding of bulk non-equilibrium effects that had been investigated extensively over the last years.[22-25]

It is worth mentioning that most of the studies in the literature concern steady state interfacial properties of layers formed by polymer – surfactant mixtures. Although the observed behavior may be far from equilibrium, technological and industrial problems are dominated by systems that rely on the response of the interface to mechanical perturbations, i.e. their rheological response.[26-28] The seminal works on the interfacial rheological characterization of polyelectrolyte – charged surfactants were carried out by Regismond et al.,[29-30] and showed the existence of a synergetic effect of the polymer – surfactant interaction on the adsorption at fluid interfaces of different mixtures. Later, Monteux et al.[31] tried to correlate the rheological response, both shear and dilational, to the stability of foams in mixtures of poly(4-styrenesulfonate of sodium) and dodecyltrimethylammonium bromide at the water / vapor interface, and found that the adsorption layers showed a behavior typical of interfacial gels, inhibiting bubble coalescence and foam drainage. This physical picture was in agreement with studies of Bhattacharyya et al.[32] More recently, Noskov et al.[33-36] have studied the response upon dilation of several mixtures formed by polyelectrolytes and surfactants bearing opposite charges. They concluded that in most cases the interfacial response upon mechanical deformation depends on the heterogeneity of the formed layers. It is important to recall that most of the aforementioned studies considered dilute mixtures, which present a composition that is far from consumer products. Consumer products contain polyelectrolyte concentrations well above the overlapping concentration $c*$[37-38], i.e. the polymer concentration above which the physico-chemical properties of the polyelectrolyte lose their dependence on the molecular weight, assuming for PDADMAC a value about 0.4 g/L,[39] and it is known that the properties of diluted and concentrated polymer systems are different.[40] This situation was also found to be the case in our previous works.[41-42] Thus, it is expected that the dynamics of interfaces may be also affected by the nature of the solutions.

The present work is devoted on the study of the interfacial properties, both equilibrium and dynamics, of layers formed by poly(diallyldimethylammonium chloride) (PDADMAC) and sodium N-lauroyl-N-methyltaurate (SLMT). The interest in studies of this polyelectrolyte – surfactant mixture is associated with their potential industrial interest for their application in cosmetics. SLMT presents similar performance for cleansing and conditioning purposes to sodium dodecyl sulfate (SDS), which has been for long time the preferred surfactant used in the fabrication of shampoos and hair care products, but SLMT is more friendly for human and environmental health.[41, 43] Furthermore, SLMT is less susceptible to degradation than SDS.[44]

The adsorption kinetics at the water / vapor interface has been measured in terms of the dynamic surface tension using a drop profile tensiometer and the dynamic surface excess by neutron reflectometry. Furthermore, the response against mechanical perturbations has been studied using a drop profile tensiometer. Even though it would be desirable to correlate the complex experimental findings with the physical phenomena present in this type of mixtures, there is no theory available for mixtures of a polymer and a surfactant yet. It is expected that this work may help to shed light on the complex physico-chemical behavior of these systems.

## 2. Experimental Section

**2.1 Chemicals.** Ultrapure deionized water used for cleaning and solution preparation had a resistivity higher than 18 MΩ·cm and a total organic content lower than 6 ppm (Younglin 370 Series, South Korea). PDADMAC was purchased from Sigma-Aldrich (Germany), and had an average molecular weight in the 100 – 200 kDa range, and was used without further purification. SLMT and perdeuterated SLMT were synthetized and purified following the procedures described in Ref. [42]. For neutron reflectometry experiments, deuterated water ($D_2O$, isotopic purity > 99.9 wt%) was purchased from Sigma-Aldrich (Germany).

The pH of all solutions was adjusted to 5.6 using glacial acetic acid (purity > 99 %) and the ionic strength was kept constant by adding 0.3 wt% of KCl (purity > 99.9 %). The PDADMAC concentration was kept constant at 0.5 wt% in all the samples.

The PDADMAC – SLMT mixtures were prepared by weight following the procedure described in our previous work.[42]

## 2.2 Techniques

*a. Surface tension measurements.* The dynamic and equilibrium surface tensions ($\gamma$) of SLMT solutions and PDADMAC – SLMT mixtures were measured using a Profile Analysis Tensiometer (PAT1-Sinterface, Germany).[45-47] This method relies on the acquisition of the profile of a solution drop formed at the tip of a steel capillary. The $\gamma$ value is obtained by fitting the acquired profile of the axis-symmetric drop using the Laplace equation. The adsorption at the water / vapor interface was measured until steady state was reached. Special care was taken to minimize evaporation effects as was detailed in our previous work.[42]

The drop profile analysis tensiometer also allows obtaining information on the response of the interface upon dilation, providing information on the frequency dependences of the complex viscoelastic modulus in the 5 – 200 mHz range. All experiments were carried out at 25.0 ± 0.1°C. The surface pressure is defined as $\Pi(c) = \gamma_0 - \gamma(c)$, where $\gamma_0$ is the surface tension of the bare water / vapor interface and $\gamma(c)$ is the surface tension of the solution/vapor interface.

*b. Neutron reflectometry.* Neutron reflectometry experiments were carried out using the time-of-flight horizontal reflectometer FIGARO at the Institut Laue-Langevin (ILL, Grenoble, France).[48] Neutron reflectivity profiles of $\log_{10} R(Q)$ were recorded with neutrons of wavelengths $\lambda$ = 2 – 30 Å at an incident angle $\theta$ = 0.622°, where R is the reflectivity and $Q = \dfrac{4\pi}{\lambda}\sin\theta$ is the momentum transfer. The analysis was carried out using Motofit.[49]

Mixtures of PDADMAC with deuterated SLMT were measured in air contrast matched water, which is a mixture of 8.1% v/v $D_2O$ in $H_2O$ that has zero scattering length density.[17] A direct measure of the surfactant surface excess in the mixture was performed by neglecting the minimal scattering contribution of the polyelectrolyte. Additional details of the experimental procedure and calculations used can be found in Ref. [42].

## 3. Theory

Thermodynamics models for describing the adsorption of surface active species at the water / vapor interface can help to the better understanding of the physico – chemical bases underlying the adsorption process. For simple surfactants, the Frumkin model is one of the most extended models. It reads as follows[50]

$$bc = \frac{\Gamma \omega}{1 - \Gamma \omega} e^{-2a\Gamma \omega} \qquad (1)$$

$$-\frac{\Pi \omega}{RT} = \ln(1 - \Gamma \omega) + a(\Gamma \omega)^2 \qquad (2)$$

where c and $\Gamma$ are the bulk concentration and the surface excess, respectively, $\omega$ represents the area per molecule at the saturated interface, b corresponds to the adsorption equilibrium constant, $a$ is the parameter of molecular interaction, and R and T are the gas constant and the absolute temperature, respectively.

The theoretical description of the experimental results for polymer – surfactant mixtures on the basis of traditional thermodynamics models is difficult, mainly due to the particular characteristic of such systems as will be discussed below. Therefore, only an empirical description of the experimental findings reminiscent of a theoretical model describing the adsorption of molecules in two different states at the interface can be used to give a qualitative analysis of the complex phenomenology associated with the adsorption of

polymer – surfactant mixtures at the water / vapor interface.[48-49] The set of empirical equations are summarized as follows,

$$bc = \frac{\Gamma_1 \omega}{(1-\Gamma\omega)^{\omega_1/\omega}} \tag{3}$$

$$-\frac{\Pi\omega}{RT} = \ln(1-\Gamma\omega) \tag{4}$$

with $\Gamma_i$ (i = 1, 2) and $\omega_i$ (i = 1, 2) being the surface excesses and the area per molecule for molecules in different states, and $b = b_1$ is the adsorption equilibrium constant in state 1. The total surface excess $\Gamma$ and the mean area per molecule $\omega$ can be defined as

$$\Gamma = \Gamma_1 + \Gamma_2 \tag{5}$$

$$\omega\Gamma = \theta = \omega_1\Gamma_1 + \omega_2\Gamma_2 \tag{6}$$

where $\theta$ represents the surface coverage. The ratio of adsorbed molecules in the two different states is given by

$$\frac{\Gamma_2}{\Gamma_1} = \exp\left(\frac{\omega_2 - \omega_1}{\omega}\right)\left(\frac{\omega_2}{\omega_1}\right)^\alpha \exp\left(-\frac{\Pi(\omega_2 - \omega_1)}{RT}\right) \tag{7}$$

where α is a constant accounting for the potential different adsorption of the species in the state 2 than in state 1. Furthermore, it is necessary to consider the role of the two-dimensional compressibility ε of the molecules in state 1,

$$\omega_1 = \omega_{10}(1 - \varepsilon\Pi\theta) \tag{8}$$

where $\omega_{10}$ is the area per molecule in the state 1 at $\Pi$ = 0. For the scope of this study, the empirical equations try to stress the complexity of the adsorption process of polymer – surfactant mixtures. Thus, the reported parameter must be considered only as a set of fitting parameters providing predictions for the concentration dependences of the surface pressure Π, the surface excess Γ and the viscoelastic moduli $\varepsilon^*$. Such predictions can be compared with the experimental results obtained by tensiometry and neutron reflectometry.

Moreover, the adsorption models allow one to obtain a qualitative prediction on the adsorption kinetics at the interface as the the temporal dependences of the surface tension and of the surface excess for a specific concentration of surface active specie. For this purpose, the dependences of the surface excesses extracted from the analysis of the surface tension isotherms must be combined with the Ward-Tordai equation,[51]

$$\Gamma(t) = 2\sqrt{\frac{D}{\pi}} \left[ c_0 \sqrt{t} - \int_0^{\sqrt{t}} c(0, t-t') d\sqrt{t'} \right] \qquad (9)$$

with D being the diffusion coefficient, $c_0$ the initial bulk concentration, t the time and t' a dummy integration variable. A last aspect that theoretical models are able to predict is the concentration dependences of the dilational viscoelastic moduli at fixed frequencies. This is based in the combination of the adsorption models with the definition of the interfacial dilational viscoelastic moduli as the increase of surface tension against a small change of the interfacial area A

$$\varepsilon^* = \frac{d\gamma}{d \ln A} \qquad (10)$$

## 4. Results and discussion

### 4.1. Equilibrium surface tension

Figure 1a shows the dependence of the surface pressure on the bulk surfactant concentration for SLMT solutions and mixtures of PDADMAC and SLMT, as obtained using the drop profile tensiometer.

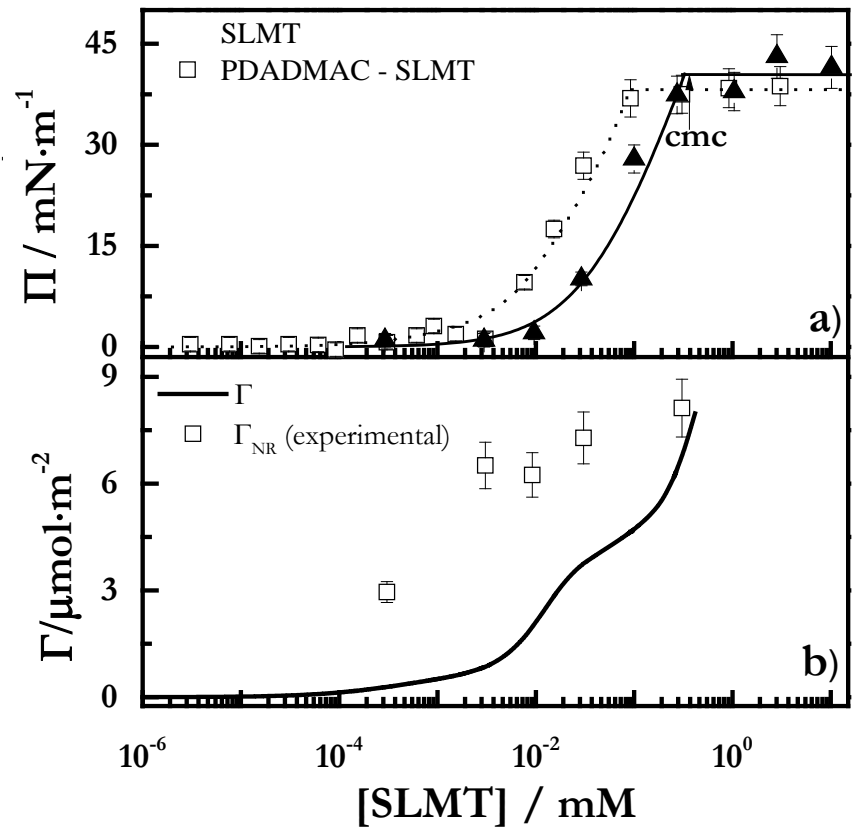

Figure 1. (a) Surface pressure isotherm for SLMT and PDADMAC – SLMT mixtures obtained by the drop profile tensiometer. Symbols represent the experimental results and the lines correspond to the values calculated with the parameters in Table 1. The dotted line corresponds to PDADMAC – SLMT and the continuous line to SLMT. (b) Total surface excess for PDADMAC – SLMT mixtures ($\Gamma$) calculated using the parameters is Table 1 and experimental total surface excess ($\Gamma_{NR}$) obtained from neutron reflectometry measurements.

SLMT exhibits an initial steep increase of the surface pressure with increasing bulk surfactant concentration, as expected for surfactant solutions, until the critical micellar concentration (cmc) is reached at around $10^{-1}$ mM. When the adsorption of PDADMAC – SLMT mixtures is considered, a synergetic effect was found between the polyelectrolyte and the surfactant on the surface pressure, i.e. a shift of the surface pressure occurs for surfactant concentrations by about one order of magnitude to lower bulk surfactant concentrations. The experimental results of pure SLMT are well described by the Frumkin isotherm.[50, 52-55] Table 1 summarizes the parameters obtained from the fit of the experimental results.

Table 1. Parameters of the calculated isotherms that described the interfacial adsorption of SLMT (Frumkin isotherm) and mixtures of PDADMAC – SLMT (Empirical isotherm).

| SLMT | PDADMAC – SLMT mixtures |
|---|---|
| $10^{-5}\,\omega(m^2/mol) = 1.27$ | $10^{-5}\,\omega_{10}(m^2/mol) = 2.70$ |
| $a = 1.10$ | $10^{-6}\,\omega_2(m^2/mol) = 1.90$ |
| $10^{-1}\,b\,(l/mmol) = 1.07$ | $10^{-2}\,b\,(l/mmol) = 2.57$ |
|  | $\alpha = 0.718$ |
|  | $10^3\,\varepsilon\,(m/mN) = 6.402$ |

For the polyelectrolyte – surfactant mixtures, the situation is more complex due to their complex character that requires additional considerations on the development of theoretical models to provide accurate description of the experimental results. It is possible to find in the literature theoretical models describing the adsorption of protein – surfactant mixtures.[52, 56-57]. Despite the binding isotherm has shown that the free surfactant concentration is almost zero,[42] which might allow one to approximate the system as a quasi-binary mixture, the following reasons make the theory non-applicable to the present mixture: a) most proteins are surface active compounds whereas PDADMAC has a negligible surface activity. b) In the system studied, the polymer concentration is above the overlapping concentration, which does not fulfill some of the conditions implicit in the theoretical model.[58] Furthermore, the interface contains both polymer + surfactant complexes and probably free surfactant (see section on interfacial rheology).

Due to the above reasons, we have limited ourselves to the application of an empirical fitting of $\gamma_{eq}$, $\gamma(t)$ using a set of equation equation with concentration-independent parameters, yet which also allows us to calculate the total surface excess and the dilational elasticity. We stress that, despite the fact that the equations derive from a theoretical model,[59] no physical meaning can be ascribed to the values obtained for the parameters given in Table 1. Nevertheless, successful application of the empirical fitting to the experimental data would represent a useful indication of the complexity of the formation of the interfacial layer. Figure 1a shows that the calculated curve is able to predict, at least qualitatively, the concentration dependence of the surface tension with the parameters given in Table 1. Figure 1b shows the surface excess curve calculated from the parameters for the PDADMAC – SLMT mixtures at the water / vapor interface. The calculated curve evidences a clear increase of the total surface excess with the bulk concentration in accordance with the dependence found for the total surface excess obtained by neutron reflectometry using structural analysis, as discussed in our previous work (see Figure 1b).[42]

The calculated surface concentration is able to provide a qualitative explanation for the dependence of the total surface excess on the surfactant concentration, although the experimental values are more than twice the theoretical ones. The discrepancies are due to the differences existing between the physico-chemical picture described by the reorientation model and the real situation of adsorbing polyelectrolyte – surfactant mixtures.

## 4.2. Adsorption kinetics

Further insight in the adsorption mechanism of PDADMAC – SLMT mixtures at the water / vapor interface can be obtained from the analysis of the time dependences of the surface tension obtained using a drop profile tensiometer. Figure 2 shows the time dependence of surface tension for the adsorption of SLMT (Figure 2a) and PDADMAC – SLMT mixtures (Figure 3b) at the water / vapor interface for different SLMT concentrations.

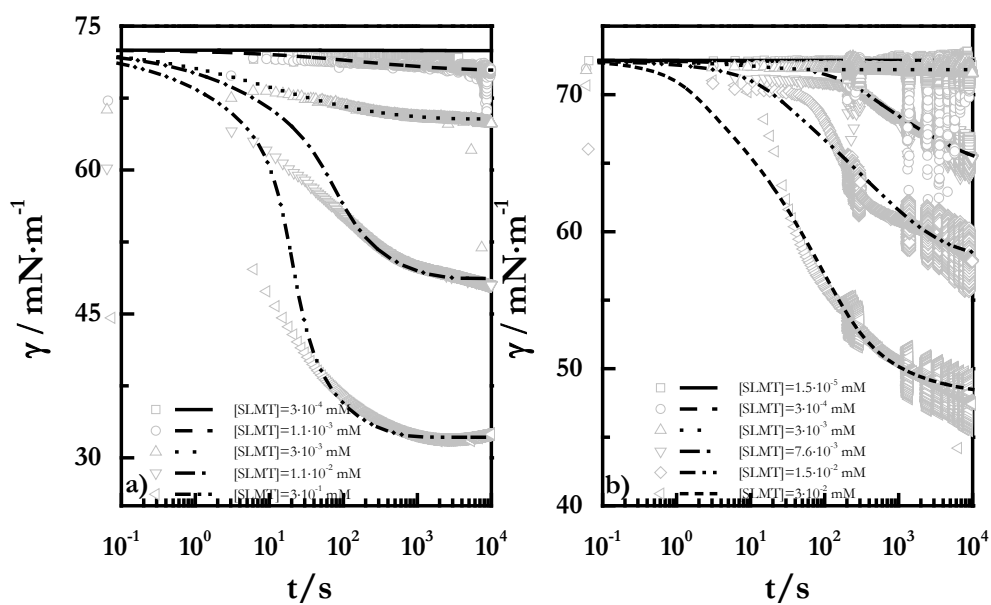

Figure 2. (a) Dynamic surface tension for SLMT solutions with concentrations below the cmc. The symbols represent the experimental results obtained using a drop profile tensiometer and the lines correspond to the fitted model using the parameters obtained for the Frumkin isotherm reported in Table 1. (b) Dynamic surface tension for PDADMAC – SLMT solutions. The symbols represent the experimental results obtained using a drop profile tensiometer and the lines correspond to the calculated curves using the parameters reported in Table 1. Note that in the data corresponding to the PDADMAC – SLMT surface tension oscillations are shown. These oscillations correspond to sinusoidal surface area perturbations

applied to the solution drops during the adsorption process to evaluate the evolution of the dilational elastic modulus during the adsorption process (see below).

The relatively long time needed for the polyelectrolyte – surfactant layer equilibration, almost three hours, for concentrations close to the cmc of the pure surfactant contrasts with the lower equilibration time needed for the pure surfactant layers, below one hour and with an important dependence on the surfactant concentration. In order to obtain further insights in the adsorption behavior, the experimental data were analyzed introducing the dependences on the surface excesses extracted from the analysis of the surface tension isotherm via the Ward-Tordai equation defined in Eq. (9).[51]

Figure 2a presents the experimental values of the dynamic surface tension for SLMT solution and the theoretical curves obtained from the Ward-Tordai equation, using a fix value for the diffusion coefficient of about $10^{-10}$ m$^2$/s for all the surfactant concentrations,[52-53] and the parameters of the Frumkin model reported in Table 1. The surface tension for the adsorption of SLMT at the water / vapor interface as expected decreases with time. This is the faster the higher the bulk surfactant concentration is. The theoretical and experimental curves present a qualitatively good agreement, especially for times above $10^2$ s. However, it is worth mentioning that the qualitative agreement of the experimental data and the theoretical curves allows one to confirm the relatively good description provided by the Frumkin model for the adsorption of SLMT.

For the PDADMAC – SLMT mixtures, it was considered that it is possible to use a procedure for the analysis of the experimental results similar to that used for the surfactant solution. However, in this case the adsorption equilibrium has been fitted using Eqs. (3) – (8). Figure 2b shows the experimental data obtained using a drop profile tensiometer for the dynamic surface tension of the mixed solutions and the calculated curves obtained using the parameters in Table 1.[52-53]

The qualitative trend of the experimental results is similar for both the mixture and the pure surfactant. However, the agreement of the curves obtained by fitting of the experimental results of dynamic surface tension is better for the mixtures. In that case, discrepancies between the experimental results and the calculated ones appear again with the increase of the SLMT concentration. However, for PDADMAC – SLMT mixtures the calculated curves show a faster decrease than the experimental ones. However, similarly to that found for the pure surfactant the theoretical curves provide a good description of the long-time tail of the dynamic surface tension curves.

A more accurate description of the adsorption of the mixtures can be carried out in terms of the time dependence of the surfactant surface excess at the water / vapor interface obtained by direct measurements using neutron reflectometry.[42] Figure 3 represents the time dependence of the surfactant surface excess, $\Gamma_{SLMT}$, for the adsorption of polyelectrolyte – surfactant mixtures at different SLMT concentrations (note that the technique is not sensitive to the present of polyelectrolyte at the interface). Notice that the surface excesses in Figure 3 correspond only to the surfactant at the interface, whereas in Figure 1b it was showed the total surface excesses which include the polyelectrolyte and surfactant amounts at the interface.

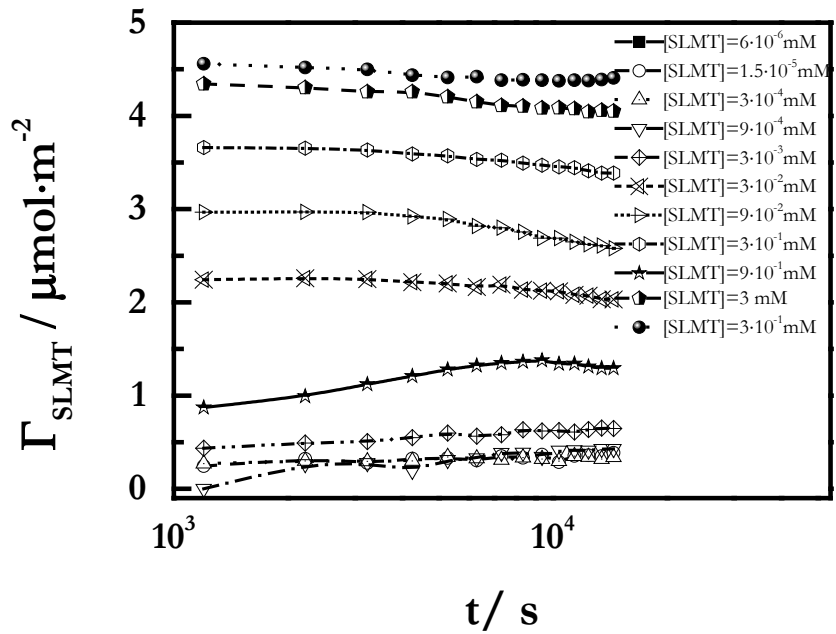

Figure 3. Dynamic surfactant surface excesses for the adsorption of PDADMAC – SLMT solutions with different concentrations at the water / vapor interface as obtained using neutron reflectometry. Lines are guides for the eyes. Notice that the error bar associated with each individual point is below the 5% of its value.

The $\Gamma_{SLMT}$ dependences on time show the existence of an adsorption of almost 80 % of the equilibrium surface excess during the first $10^3$ s, which may be explained assuming that the first step of the adsorption process of the mixtures occurs by diffusion from the bulk. From, the results, it is possible to assume that the reorganization of the material at the interface allows either further adsorption at the lowest surfactant concentration, or the squeezing out of material from the interface, leading to a decrease of $\Gamma_{SLMT}$ for the highest surfactant concentration. It is worth mentioning that for the lowest SLMT concentrations there is adsorption of material, as evidenced by the change of $\Gamma_{SLMT}$, without any appreciable change of the surface tension. Thus, it is possible to assume that for the lowest concentration, the surface tension does not provide a real evaluation of the thermodynamic equilibrium of the interfacial layer, as was previously discussed for different systems in literature.[33, 60-61]

### 4.3. Dilational rheology

The above discussion concerns features of the adsorption layers and the adsorption kinetics for samples of fixed surface area. However, many technological applications are associated with the response of the system upon mechanical perturbations. Therefore, an understanding of the relaxation mechanisms appearing in the interfacial layers upon a dilational perturbation provide additional useful information on the physico-chemical properties of the system, which is relevant for industrial applications, including foam stabilization or thin film deposition.[14,62-66] The dependences of the dilational viscoelasticity moduli $\varepsilon^* = \varepsilon' + i\varepsilon''$ ($\varepsilon'$ is referred to the dilational elastic modulus and $\varepsilon''$ is the viscous modulus) on the deformation frequency and the interfacial concentration,[67] will be discussed in this section. In order to complement the above discussion of the adsorption kinetics, the evolution of the elastic modulus during the adsorption processes will be analyzed at a fixed frequency of 50 mHz. It is worth mentioning that the viscous modulus presents a negligible value for all the studied systems, therefore for the sake of simplicity no discussion on its dependences will be here included, note that $\varepsilon'/\varepsilon''$ becomes higher as the surfactant concentration increases. Figure 4 shows the dependence of the elastic modulus over time for the adsorption of PDADMAC – SLMT solutions at the water / vapor interface.

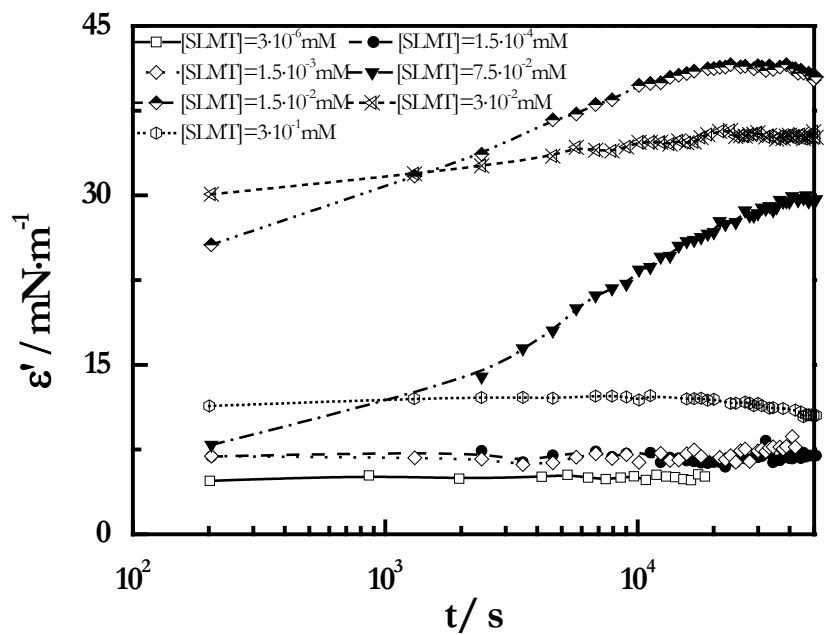

Figure 4. Time dependences for the elastic modulus obtained by the oscillating drop method at a deformation frequency of 50 mHz for PDADMAC – SLMT solutions with different surfactant concentrations. The lines are guides for the eyes.

The time dependence of the elastic modulus agrees well with the results above discussed for the dynamic surface tension and the dynamic surface excess. The most significant change in the layer properties occur within the first stage of adsorption (about $10^3$ s) for almost all the concentrations considered. The results show that for the lowest surfactant concentration, a non-negligible elastic modulus is found which is reasonable considering the non-negligible adsorption of material obtained from neutron scattering. The elastic modulus increases with surfactant concentration in the region of higher decrease of the surface tension (see Figure 1a), which is explained by the adsorption of the most part of material at the interface, in this region the change of the elastic modulus with time is slower. Then, a decrease of the elastic modulus with concentration until values close to water is observed when the interface is saturated with surfactant. The dynamic elastic modulus provides complementary information to the above discussion, more specifically about the different processes occurring at different time scales during the equilibration of the interfacial layers. Figure 5 shows the results, using a drop profile tensiometer, obtained for solutions of two different concentrations of SLMT and PDADMAC – SLMT mixtures.

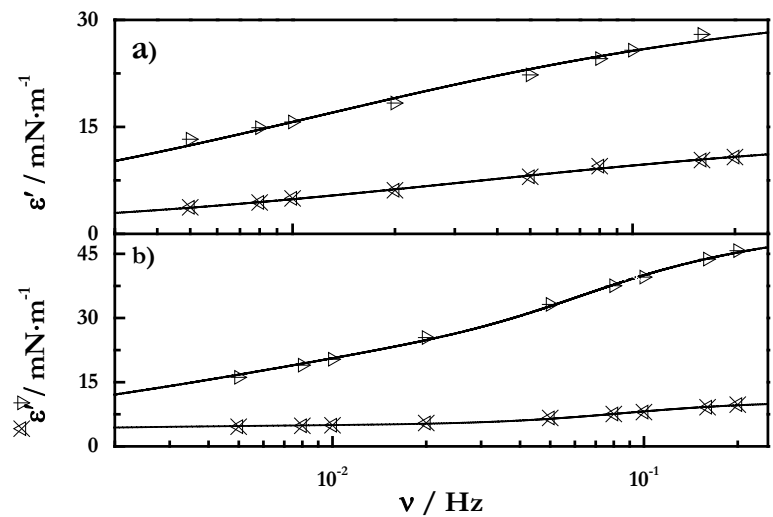

Figure 5. Frequency dependences of the elastic moduli for solutions of SLMT (a) and PDADMAC – SLMT mixtures (b). In both panels, the symbols are referred to different surfactant concentrations ($\ast$ = 0.03 mM and $\triangleright$ = 0.09 mM) and the lines represent the best fit of the experimental curves to the corresponding theoretical models described by Eq. (11) for pure surfactant and Eq. (12) for PDADMAC – SLMT mixtures (see Figure 6).

A detailed analysis of the frequency dependences evidences different behavior for surfactant solutions and for the polyelectrolyte – surfactant mixtures. The relaxation of surfactant layers (Figure 5a) at the water / vapor interface is easily explained assuming the existence of a diffusion controlled process as described by the model proposed by Lucassen – van den Tempel,[68-70] providing the following description of the frequency dependence of the viscoelasticity modulus

$$\varepsilon^* = \frac{1+\xi+i\xi}{1+2\xi+2\xi^2} \qquad (11)$$

where $\xi = \sqrt{\frac{\nu_D}{\nu}}$ with $\nu_D$ being the characteristic frequency of the diffusion exchange and $\nu$ is the frequency of deformation. Independently of the concentration, the characteristic frequency of the diffusion, $\nu_D$, assumes values in the range $10^{-2} - 10^{-1}$ Hz.

In the case of PDADMAC – SLMT mixtures (Figure 6b), the diffusion controlled mechanism alone does not account for the internal relaxation in the interfacial layers. Similar results were also obtained for simpler polymer solutions.[71] Thus, the introduction of another relaxation process to the expression accounting for the viscoelastic modulus is needed. For this purpose, we follow the description provided by Ravera et al.[60, 72-73]

$$\varepsilon^* = \frac{1+\xi+i\xi}{1+2\xi+2\xi^2}\left[\varepsilon_0 + (\varepsilon_1 - \varepsilon_0)\frac{1+i\lambda}{1+\lambda^2}\right] \qquad (12)$$

where $\lambda = \nu_1/\nu$ with $\nu_1$ being the characteristic frequency of the extra relaxation process, and $\varepsilon_0$ and $\varepsilon_1$ the Gibbs elasticity and the high limit elasticity within the frequency range considered, respectively. The fact that the aforementioned model provides a description of the response upon dilation of PDADMAC – SLMT layers may be correlated with the complexity of the adsorption process discussed above on the basis of the adsorption isotherm. Combining the adsorption isotherm and the frequency dependences of the elastic modulus, it is possible to provide an explanation for the equilibration of the adsorption layers of PDADMAC – SLMT mixtures. First the diffusion-controlled adsorption of the complexes at the subsurface interface takes place, which is followed by a second process of adsorption controlled by the reorganization of the material adsorbed during the first step. This scenario is similar to that proposed by Noskov et al.[33] for the adsorption of PDADMAC – SDS at the water / vapor interface. Figure 6 shows the concentration dependences of the characteristic frequencies corresponding to the two steps involved in the equilibration of the adsorption layers of the PDADMAC – SLMT mixtures.

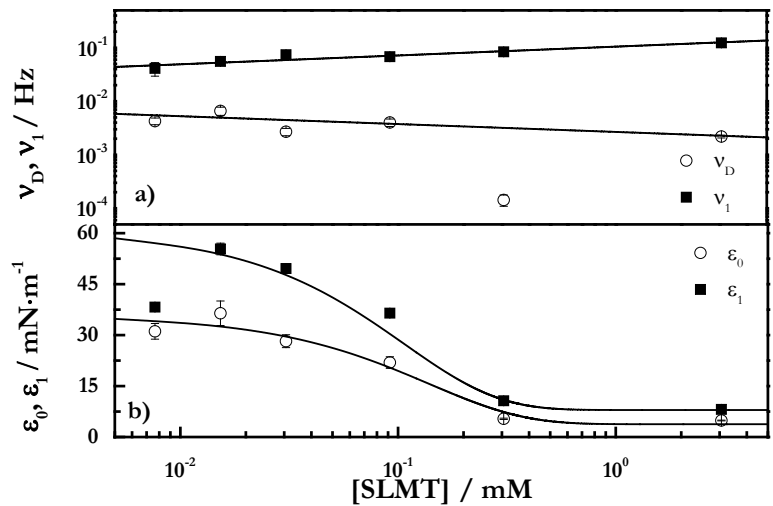

Figure 6. Concentration dependences for PDADMAC – SLMT mixtures: (a) Frequencies of the two relaxation processes. (b) Limiting elasticities. The symbols represent the experimental data and the lines are guides for the eyes. Notice that the error bars are within the size of the points.

As expected, the frequency corresponding to the interfacial relaxation process, $\nu_1$, is higher than that associated with the diffusional transport, $\nu_D$. This is easily rationalized considering that such a process occurs only once the initial adsorption of material at the fluid interface has taken place. Furthermore, the dependences of the characteristic frequencies on the SLMT concentration are different for $\nu_D$ and $\nu_1$, whereas the former is rather constant or decreases slightly as the surfactant concentration increases, $\nu_1$ increases with the SLMT concentration. This can be explained considering that the first step is slightly slowed down with the surfactant concentration because the increase of the adsorbed amount at the interface introduces a hindrance, steric or electrostatic, for the adsorption of further material, thus the first step occurs slowly. The characteristic frequency associated with the diffusion (in the range $10^{-2} - 10^{-3}$ Hz) is similar to that associated with the diffusion of supramolecular species found in a previous study.[60] The detailed analysis of the limiting elasticity values (Figure 6b) shows clearly that both $\varepsilon_0$ and $\varepsilon_1$ decrease with the surfactant concentration, being as it is expected $\varepsilon_1 > \varepsilon_0$.

Figure 7a shows the concentration dependences of the elastic modulus of SLMT layers. For sake of simplicity only the results corresponding to some of the frequencies studied are shown. Other frequencies show similar dependences.

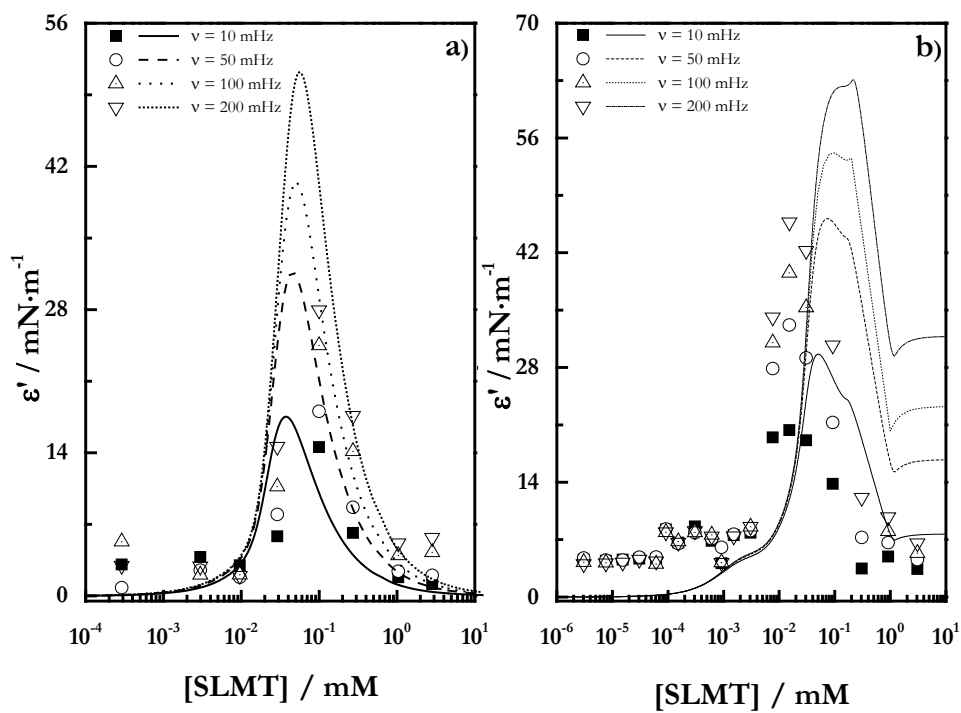

Figure 7. (a) Concentration dependences of the elastic modulus for SLMT adsorption layers as were obtained from oscillating drop experiments. (b) Concentration dependences of the elastic modulus for PDADMAC – SLMT adsorption layers as were obtained from oscillating drop experiments. In both panels, the symbols represent the experimental results and the lines represent the values calculated with the parameters in Table 1.

The results show an increase of the elastic modulus from quasi-null values to reach a maximum which depends on the deformation frequencies, and then further increases of the surfactant concentration leads again to the decrease of the elastic modulus to reach values close to zero. This is the typical behavior expected for the surfactant concentration dependence of the elastic modulus. The theoretical curves calculated using the parameters of the Frumkin isotherm showed in Table 1 agrees qualitatively with the experimental curves, especially as the concentration gets closer to the cmc of the surfactant and at larger frequency.

For PDADMAC – SLMT mixtures, the dependence of the elasticity on the surfactant concentration is similar to that found for pure surfactant, as it is shown in Figure 7b. The comparison of the calculated and the experimental data evidences that Eqs. (3) to (8) do not provide a good description of the dilational elastic modulus, which follows given the complexity of the mixture and that the calculated values of $\Gamma$ are significantly smaller than the experimental ones.

## 5. Conclusions

This work has focused on the adsorption process from mixtures formed by a polyelectrolyte PDADMAC and an oppositely charged surfactant SLMT at the water / vapor interface. The system was studied using a combination of surface tension measurements and neutron reflectometry. The experimental results have evidenced that the adsorption of PDADMAC – SLMT mixtures is rather different from that found for the adsorption of pure SLMT solutions. The experimental results indicate that layer formation at the water / vapor interface occurs in a two-step process in which first the adsorption of the bulk formed complexes takes places, and once they are adsorbed an interfacial reorganization of the adsorbed material occurs until the equilibration of the adsorbed layer. This behavior has been described qualitatively in terms of a set of empirical equations, which is compatible with the appeareance of two relaxation process in dynamic measurements. Despite the fact that such an equation is able to predict the tensiometric results, it does not provide an accurate quantitative description neither of the surface excesses nor the dilational elastic modulus. Thus, the obtained results underline the complexity of the adsorption processes involved in the adsorption behavior of polyelectrolyte – surfactant mixtures at fluid interfaces and allow us to conclude that such complex behavior cannot be formally described on the bases of simple thermodynamics models. Therefore, a deeper study on the adsorption of such mixtures must be carried out in order to establish accurately an explanation of the adsorption mechanism, and to develop a theoretical description of this type of mixtures, which is urgently required for the design and optimization of the next generation of polyelectrolyte – surfactant mixtures of technology interest.

**Acknowledgements**

This work was funded in part by MINECO under grants FIS 2014-62005-EXP and CTQ-2016- 78895-R, by EU under Marie Curie ITN CoWet (Grant Number 607861). This work is based upon work from COST Action MP-1305. The authors are grateful to the CAI of Espectroscopia y Correlación of UCM for the use of their facilities and to the ILL for an allocation of neutron beam time on FIGARO as well as the Partnership for Soft Condensed Matter for access to ancillary equipment during the experiment.

# References

1. Piculell, L.; Lindman, B., Association and Segregation in Aqueos Polymer/Polymer, Polymer/Surfactant, and Surfactant/Surfactant Mixtures: Similarities and Differences. *Adv. Colloid Interface Sci.* **1992,** *41*, 149-178. doi: 10.1016/0001-8686(92)80011-l.
2. Llamas, S.; Guzmán, E.; Ortega, F.; Baghdadli, N.; Cazeneuve, C.; Rubio, R. G.; Luengo, G. S., Adsorption of polyelectrolytes and polyelectrolytes-surfactant mixtures at surfaces: a physico-chemical approach to a cosmetic challenge. *Adv. Colloid Interface Sci.* **2015,** *222*, 461-487. doi: 10.1016/j.cis.2014.05.007.
3. Bain, C. D.; Claesson, P. M.; Langevin, D.; Meszaros, R.; Nylander, T.; Stubenrauch, C.; Titmuss, S.; von Klitzing, R., Complexes of surfactants with oppositely charged polymers at surfaces and in bulk. *Adv. Colloid Interface Sci.* **2010,** *155*, 32–49. doi: 10.1016/j.cis.2010.01.007.
4. Goddard, E. D.; Ananthapadmanabhan, K. P., *Application of Polymer–Surfactant Systems*. Marcel Dekker, Inc.: New York, United States of America, 1998.
5. Szczepanowicz, K.; Bazylińska, U.; Pietkiewicz, J.; Szyk-Warszyńska, L.; Wilk, K. A.; Warszyński, P., Biocompatible long-sustained release oil-core polyelectrolyte nanocarriers: From controlling physical state and stability to biological impact. *Adv. Colloid Interface Sci.* **2015,** *222*, 678-691.. doi: 10.1016/j.cis.2014.10.005.
6. Guzmán, E.; Llamas, S.; Maestro, A.; Fernández-Peña, L.; Akanno, A.; Miller, R.; Ortega, F.; Rubio, R. G., Polymer-surfactant systems in bulk and at fluid interfaces. *Adv. Colloid Interface Sci.* **2016,** *233*, 38-64. doi: 10.1016/j.cis.2015.11.001.
7. Chiappisi, L.; Hoffmann, I.; Gradzielski, M., Complexes of oppositely charged polyelectrolytes and surfactants – recent developments in the field of biologically derived polyelectrolytes. *Soft Matter* **2013,** *9*, 3896-3909. doi: 10.1039/c3sm27698h.
8. Guzmán, E.; Chuliá-Jordán, R.; Ortega, F.; Rubio, R. G., Influence of the percentage of acetylation on the assembly of LbL multilayers of poly(acrylic acid) and chitosan. *Phys. Chem. Chem. Phys.* **2011,** *13*, 18200-18207. doi: 10.1039/c1cp21609k.
9. Nylander, T.; Samoshina, Y.; Lindman, B., Formation of polyelectrolyte–surfactant complexes on surfaces. *Adv. Colloid Interface Sci.* **2006,** *123–126*, 105-123. doi: 10.1021/acs.langmuir.7b01288.
10. Thalberg, K.; Lindman, B., Polymer-surfactant interactions recent developments. In *Interactions of Surfactants with Polymers and Proteins*, Goddard, E. D.; Ananthapadmanabhan, K. P., Eds. CRC Press: Boca Raton, United States of America, 1993.

11. Von Klitzing, R., Internal Structure of polyelectrolyte multilayer assemblies. *Phys. Chem. Chem. Phys.* **2006,** *8*, 5012-5033. doi: 10.1039/b607760a.

12. Ferreira, G. A.; Loh, W., Liquid crystalline nanoparticles formed by oppositely charged surfactant-polyelectrolyte complexes. *Curr. Opin. Colloid Interface Sci.* **2017,** *32*, 11-22. doi: 10.1016/j.cocis.2017.08.003.

13. Goddard, E. D.; Hannan, R. B., Cationic polymer/anionic surfactant interactions *J. Colloid Interface Sci.* **1976,** *55*, 73-79. doi: 10.1016/0021-9797(76)90010-2.

14. Bergeron, V.; Langevin, D.; Asnacios, A., Thin-film forces in foam films containing anionic polyelectrolyte and charged surfactants. *Langmuir* **1996** *12*, 1550-1556. doi: 10.1016/0021-9797(76)90010-2.

15. Ritacco, H.; Albouy, P.-A.; Bhattacharyya, A.; Langevin, D., Influence of the polymer backbone rigidity on polyelectrolytesurfactant complexes at the air/water interface. *Phys. Chem. Chem. Phys.* **2000,** *2*, 5243−5251. doi: 10.1039/b004657o.

16. Stubenrauch, C.; Albouy, P.-A.; von Klitzing, R.; Langevin, D., Polymer/surfactant complexes at the water/air interface: a surface tension and x-ray reflectivity study. *Langmuir* **2000,** *16*, 3206-3213. doi: 10.1016/0021-9797(76)90010-2.

17. Lu, J. R.; Thomas, R. K.; Penfold, J., Surfactant layers at the air/water interface: structure and composition. *Adv. Colloid Interface Sci.* **2000,** *84*, 143-304. doi: 10.1016/S0001-8686(99)00019-6.

18. Narayanan, T.; Wacklin, H.; Konovalov, O.; Lund, R., Recent applications of synchrotron radiation and neutrons in the study of soft matter *Crystallography Rev.* **2017** *23*, 160-226. doi: 10.1080/0889311x.2016.1277212.

19. Braun, L.; Uhlig, M.; Von Klitzing, R.; Campbell, R. A., Polymers and surfactants at fluid interfaces studied with specular neutron reflectometry. *Adv. Colloid Interface Sci.* **2017,** *247*, 130-148. doi: 10.1016/j.cis.2017.07.005.

20. Varga, I.; Campbell, R. A., General Physical Description of the Behavior of Oppositely Charged Polyelectrolyte/Surfactant Mixtures at the Air/Water Interface. *Langmuir* **2017,** *33*, 5915-5924. doi: 10.1021/acs.langmuir.7b01288.

21. Campbell, R. A.; Arteta, M. Y.; Angus-Smyth, A.; Nylander, T.; Noskov, B. A.; Varga, I., Direct Impact of Non-Equilibrium Aggregates on the Structure and Morphology of Pdadmac/SDS Layers at the Air/Water Interface. *Langmuir* **2014,** *30*, 8664-8774. doi: 10.1021/la500621t.

22. Kogej, K.; J., S., Surfactant Binding to Polyelectrolytes. In *Surfactant Science Series*, Radeva, T., Ed. Marcell Dekker Inc.: New York, United States of America, 2001; Vol. 99, pp 793-827.

23. Li, D.; Kelkar, M. S.; Wagner, N. J., Phase Behavior and Molecular Thermodynamics of Coacervation in Oppositely Charged Polyelectrolyte/Surfactant Systems: A Cationic Polymer JR 400 and Anionic Surfactant SDS Mixture. *Langmuir* **2012,** *28*, 10348−10362. doi: 10.1021/la301475s.

24. Hansson, P.; Lindman, B., Surfactant-polymer interactions. *Curr. Opin. Colloid Interface Sci.* **1996,** *1*, 604-613. doi: 10.1016/s1359-0294(96)80098-7.

25. Bergfeldt, K.; Piculell, L.; Linse, P., Segregation and Association in Mixed Polymer Solutions from Flory-Huggins Model Calculations. *J. Phys. Chem.* **1996,** *100*, 3680-3687. doi: 10.1021/jp952349s.

26. Fuller, G. G.; Vermant, J., Complex Fluid-Fluid Interfaces: Rheology and Structure. *Annu. Rev. Chem. Biomol. Eng.* **2012,** *3*, 519-543. doi: 10.1146/annurev-chembioeng-061010-114202.

27. Angus-Smyth, A.; Campbell, R. A.; Bain, C. D., Dynamic Adsorption of Weakly Interacting Polymer/Surfactant Mixtures at the Air/Water Interface. *Langmuir* **2012,** *28*, 12479−12492. doi: 10.1021/la301297s.

28. Aidarova, S.; Sharipova, A.; Krägel, J.; Miller, R., Polyelectrolyte/surfactant mixtures in the bulk and at water/oil interfaces. *Adv. Colloid Interface Sci.* **2014,** *205*, 87-93. doi: 10.1016/j.cis.2013.10.007.
29. Regismond, S. T. A.; Winnik, F. M.; Goddard, E. D., Surface viscoelasticity in mixed polycation anionic surfactant systems studied by a simple test. *Colloids Surf. A* **1996,** *119*, 221 228.
30. Regismond, S. T. A.; Gracie, K. D.; Winnik, F. M.; Goddard, E. D., Polymer/Surfactant Complexes at the Air/Water Interface Detected by a Simple Measure of Surface Viscoelasticity. *Langmuir* **1997,** *13*, 5558-5562. doi: 10.1021/la9702289.
31. Monteux, C.; Fuller, G. G.; Bergeron, V., Shear and Dilational Surface Rheology of Oppositely Charged Polyelectrolyte/Surfactant Microgels Adsorbed at the Air-Water Interface. Influence on Foam Stability. *J. Phys. Chem. B* **2004,** *108*, 16473-16482. doi: 10.1021/jp047462+.
32. Bhattacharyya, A.; Monroy, F.; Langevin, D.; Argillier, J.-F., Surface Rheology and Foam Stability of Mixed Surfactant-Polyelectrolyte Solutions. *Langmuir* **2000,** *16*, 8727–8732. doi: 10.1016/0021-9797(76)90010-2.
33. Noskov, B. A.; Grigoriev, D. O.; Lin, S. Y.; Loglio, G.; Miller, R., Dynamic Surface Properties of Polyelectrolyte/Surfactant Adsorption Films at the Air/Water Interface: Poly(diallyldimethylammonium chloride) and Sodium Dodecylsulfate. *Langmuir* **2007,** *23*, 9641-9651. doi: 10.1021/la700631t.
34. Noskov, B. A., Dilational surface rheology of polymer and polymer/surfactant solutions. *Curr. Opin. Colloids Interface Sci.* **2010,** *15*, 229–236. doi:10.1016/j.cocis.2010.01.006.
35. Noskov, B. A.; Loglio, G.; Miller, R., Dilational surface visco-elasticity of polyelectrolyte/surfactant solutions: Formation of heterogeneous adsorption layers. *Adv. Colloid Interface Sci.* **2011,** *168*, 179-197. doi: 10.1016/j.cis.2011.02.010.
36. Lyadinskaya, V. V.; Bykov, A. G.; Campbell, R. A.; Varga, I.; Lin, S. Y.; Loglio, G.; Miller, R.; Noskov, B. A., Dynamic surface elasticity of mixed poly(diallyldimethylammoniumchloride)/sodium dodecyl sulfate/NaCl solutions. *Colloids Surf. A* **2014,** *460*, 3-10. doi: 10.1016/j.colsurfa.2014.01.041.
37. Dautzenberg, H., *Polyelectrolytes : formation, characterization, and application*. Hanser Publishers: New York, United States of America, 1994.
38. Dobrynin, A. V.; Rubinstein, M., Theory of polyelectrolytes in solutions and at surfaces. *Prog. Polym. Sci.* **2005,** *30*, 1049-1118. doi:10.1016/j.progpolymsci.2005.07.006.
39. Guzmán, E.; Ritacco, H.; Rubio, J. E. F.; Rubio, R. G.; Ortega, F., Salt-induced changes in the growth of polyelectrolyte layers of poly(diallyldimethylammoniumchloride) and poly(4-styrene sulfonate of sodium). *Soft Matter* **2009,** *5*, 2130-2142. doi: 10.1039/b901193e.
40. Doi, M.; Edwards, S. F., *The Theory of Polymer Dynamics*. Oxford Science Publications: Oxford, United Kingdom, 1990.
41. Llamas, S.; Guzmán, E.; Baghdadli, N.; Ortega, F.; Cazeneuve, C.; Rubio, R. G.; Luengo, G. S., Adsorption of poly(diallyldimethylammonium chloride)—sodium methyl-cocoyl-taurate complexes onto solid surfaces. *Colloids Surf. A* **2016,** *505*, 150-157. doi: 10.1016/j.colsurfa.2016.03.003.
42. Llamas, S.; Fernández-Peña, L.; Akanno, A.; Guzmán, E.; Ortega, V.; Ortega, F.; Csaky, A. G.; Campbell, R. A.; Rubio, R. G., Towards understanding the behavior of polyelectrolyte - surfactant mixtures at the water / vapor interface closer to technologically-relevant conditions. *Phys. Chem. Chem. Phys.* **2018,** *20*, doi: 10.1039/c7cp05528e.
43. Goddard, E. D.; Gruber, J. V., *Principles of Polymer Science and Technology in Cosmetics and Personal Care*. Marcel Dekker, Inc.: Basel, Switzerland, 1999.

44. Fang, J. P.; Joos, P., The Dynamic Surface Tension of SDS—Dodecanol Mixtures: 1. The Submicellar Systems. *Colloids Surf. A* **1992,** *65*, 113-120. doi: 10.1016/0166-6622(92)80266-5.

45. Wang, Y.; Kimura, K.; Dubin, P. L.; Jaeger, W., Polyelectrolyte-Micelle Coacervation: Effects of Micelle Surface Charge Density, Polymer Molecular Weight, and Polymer/Surfactant Ratio. *Macromolecules* **2000,** *33*, 3324-3331. doi: 10.1021/ma991886y.

46. Ortmann, T.; Ahrens, H.; Lawrenz, F.; Groening, A.; Nestler, P.; Guenther, J.-U.; Helm, C. A., Lipid Monolayers and Adsorbed Polyelectrolytes with Different Degrees of Polymerization. *Langmuir* **2014,** *30*, 6768-6779. doi: 10.1021/la5001478.

47. Chen, P.; Kwok, D. Y.; Prokop, R. M.; del Rio, O. I.; Sunsar, S. S.; Neumann, A. W., Axisymmetric Drop Shape Analysis (ADSA) and its applications. In *Drops and Bubbles in Interfacial Research*, Möbius, D.; Miller, R., Eds. Elsevier Science B.V.: Amsterdam, The Netherlands, 1998; Vol. 6, pp 61-139. doi: 10.1016/S1383-7303(98)80019-7.

48. Campbell, R. A.; Wacklin, H. P.; Sutton, I.; Cubitt, R.; Fragneto, G., FIGARO: the new horizontal neutron reflectometer at the ILL. *Eur. Phys. J. Plus* **2011,** *126*, 107. doi: 10.1140/epjp/i2011-11107-8.

49. Nelson, A., Co-refinement of multiple-contrast neutron/X-ray reflectivity data using MOTOFIT. *J. Appl. Crystallogr.* **2006,** *39*, 273. doi:10.1107/s0021889806005073.

50. Eisenthal, K. B., Equilibrium and dynamic processes at interfaces by second harmonic and sum frequency generation. *Annu. Rev. Phys. Chem.* **1992,** *43*, 627-661. doi: 10.1146/annurev.pc.43.100192.003211

51. Ward, A. F. H.; Tordai, L., Time Dependence of Boundary Tensions of Solutions I. The Role of Diffusion in Time Effects. *J. Chem. Phys.* **1946,** *14*, 453-461. doi: 10.1063/1.1724167.

52. Aksenenko, E. V. Adsorption Software. http://www.thomascat.info/Scientific/AdSo/AdSo.htm. (accessed September 4, 2017).

53. Aksenenko, E. V., Software tools to interpret the thermodynamics and kinetics of surfactant adsorption. In *Surfactants - Chemistry, Interfacial Properties and Application, Studies in Interface Science*, Fainerman, V. B.; Möbius, D.; Miller, R., Eds. Elsevier: Amsterdam , The Netherlands, 2001; Vol. 13, pp 619-648.

54. Zhang, R.; Somasundaran, P., Advances in adsorption of surfactants and their mixtures at solid/solution interfaces. *Adv. Colloid Interface Sci.* **2006,** *123–126*, 213-229. doi:10.1016/j.cis.2006.07.004.

55. Rosen, M. J., *Surfactants and Interfacial Phenomena*. John Wiley and Sons: Hoboken, United States of America, 2004.

56. Fainerman, V. B.; Lylyk, S. V.; Aksenenko, E. V.; Makievski, A. V.; Petkov, J. T.; Yorke, J.; Miller, R., Adsorption layer characteristics of Triton surfactants: 1. Surface tension and adsorption isotherms. *Colloids Surf. A* **2009,** *334*, 1-7. doi: 10.1016/j.colsurfa.2008.09.015.

57. Fainerman, V. B.; Lylyk, S. V.; Aksenenko, E. V.; Liggieri, L.; Makievski, A. V.; Petkov, J. T.; Yorke, J.; Miller, R., Adsorption layer characteristics of Triton surfactants 2. Dynamic surface tension and adsorption. *Colloids Surf. A* **2009,** *334*, 8-15. doi: 10.1016/j.colsurfa.2008.09.052.

58. Kotsmar, C.; Pradines, V.; Alahverdjieva, V. S.; Aksenenko, E. V.; Fainerman, V. B.; Kovalchuk, V. I.; Krägel, J.; Leser, M. E.; Noskov, B. A.; Miller, R., Thermodynamics, adsorption kinetics and rheology of mixed protein–surfactant interfacial layers. *Adv. Colloid Interface Sci.* **2009,** *150*, 41-54. doi:10.1016/j.cis.2009.05.002.

59. Fainerman, V. B.; Miller, R.; Aksenenko, E. V.; Makievski, A. V.; Krägel, J.; Loglio, G.; Liggieri, L., Effect of surfactant interfacial orientation/aggregation on adsorption dynamics. *Advances in Colloid and Interface Science* **2000,** *86* (1–2), 83-101.

60. Liggieri, L.; Santini, E.; Guzmán, E.; Maestro, A.; Ravera, F., Wide-Frequency Dilational Rheology Investigation of Mixed Silica Nanoparticle – CTAB Interfacial Layers. *Soft Matter* **2011,** *7* 6699-7709. doi: 10.1039/c1sm05257h.

61. Maestro, A.; Guzman, E.; Santini, E.; Ravera, F.; Liggieri, L.; Ortega, F.; Rubio, R. G., Wettability of silica nanoparticle–surfactant nanocomposite interfacial layers. *Soft Matter* **2012,** *8*, 837-843. doi: 10.1039/c1sm06421e.

62. Schramm, L. L., *Emulsions, Foams and Suspensions. Fundamentals and Applications*. Wiley-VCH Verlag: Weinheim, Germany, 2005.

63. Murray, B. S., Stabilization of bubbles and foams. *Curr. Opin. Colloid Interface Sci.* **2007,** *12*, 232-241. doi:10.1016/j.cocis.2007.07.009.

64. Langevin, D., Aqueous Foams: A Field of Investigation at the Frontier Between Chemistry and Physics. *ChemPhysChem* **2008,** *9*, 510 - 522. doi: 10.1016/j.cocis.2007.07.009.

65. Ropers, M. H.; Novales, B.; Boue, F.; Axelos, M. A. V., Polysaccharide/Surfactant Complexes at the Air-Water Interface: Effect of the Charge Density on Interfacial and Foaming Behaviors. *Langmuir* **2008,** *24*, 12849-12857. doi: 10.1021/la802357m.

66. Larsson, R. G., *The Structure and Rheology of Complex Fluids*. Oxford University Press: Oxford, United Kingdom, 1999.

67. Monroy, F.; Ortega, F.; Rubio, R. G., Rheology of a miscible polymer blend at the air-water interface. Quasielastic light-scattering study and analysis in terms of static and dynamic scaling laws. *J. Phys. Chem. B* **1999,** *103*, 2061-2071. doi: doi: 10.1021/la802357m.

68. Lucassen, J.; Van Den Tempel, M., Dynamic measurements of dilational properties of a liquid interface. *Chem. Eng. Sci.* **1972,** *27*, 1283-1291. doi: 10.1016/0009-2509(72)80104-0.

69. Van den Tempel, M.; Lucassen-Reynders, E., Relaxation processes at fluid interfaces. *Adv. Colloid Interface Sci.* **1983,** *18*, 281-301. doi: 10.1016/0001-8686(83)87004-3.

70. Lucassen, J., Dynamic dilational properties of composite surfaces. *Colloids Surf.* **1991,** *65*, 139- 149. doi: 10.1016/0001-8686(83)87004-3.

71. Llamas, S.; Mendoza, A. J.; Guzman, E.; Ortega, F.; Rubio, R. G., Salt effects on the air/solution interfacial properties of PEO-containing copolymers: Equilibrium, adsorption kinetics and surface rheological behavior. *J. Colloid Interface Sci.* **2013,** *400*, 49-58. doi: 10.1016/j.jcis.2013.03.015.

72. Ansari, A. A.; Kamil, M.; Kabir-ud-Din, Interaction of Oppositely Charged Polymer–Surfactant System Based on Surface Tension Measurements. *J. Petroleum Sci. Res.* **2013,** *2*, 35-40.

73. Ravera, F.; Ferrari, M.; Santini, E.; Liggieri, L., Influence of surface processes on the dilational visco-elasticity of surfactant solutions. *Adv. Colloid Interface Sci.* **2005,** *117*, 75 – 100. doi:10.1016/j.cis.2005.06.002.

**TOC Graphic**

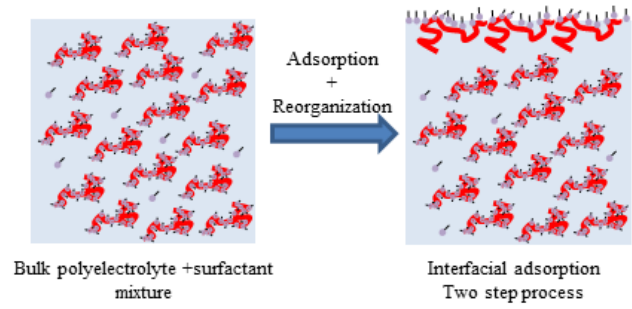